\documentclass[twocolumn,twoside]{IEEEtran} 
\usepackage{times}

\usepackage{graphicx}

\usepackage{amsmath}
\usepackage{times}
\usepackage{graphicx}
\usepackage{amsfonts}
\usepackage{amssymb}
\usepackage{cite}
\usepackage{subfigure}
\usepackage{url}
\usepackage[nolist,nohyperlinks]{acronym}
\usepackage[numbers,sort&compress]{natbib}

\DeclareGraphicsExtensions{.png,.eps,.ps,.pdf}

\usepackage{tikz}
\usetikzlibrary{arrows,	
				shapes, 
				positioning, 
				fit,	
                backgrounds, 
                mindmap,
                petri,
				intersections, 
				external 
				}
\definecolor{emerald}{RGB}{69,155,61}
\definecolor{gold}{RGB}{244,216,51}
\definecolor{pink}{RGB}{235,44,206}
\usepackage{verbatim}

\begin{acronym}
\acro{SNR}{signal-to-noise ratio}
\acrodefplural{SNR}[SNRs]{signal-to-noise ratios}
\acro{RV}{random variable}
\acrodefplural{RV}[RVs]{random variables}
\acro{LOS}{line-of-sight}
\acrodefplural{LoS}{Line-of-sight}
\acro{EH}{Energy Harvesting}
\acro{PDF}{probability density function}
\acro{CDF}{cumulative distribution function}
\acrodefplural{CDF}[CDFs]{cumulative distribution functions}
\acro{IoT}{Internet of Things}
\acro{PB}{Power Beacon}
\acrodefplural{PB}[PBs]{Power Beacons}
\acro{WPC}{Wireless Powered Communications}
\acro{WPT}{Wireless Power Transmission}
\acro{RFID}{Radio Frequency IDentification}
\acro{MC}{Monte Carlo}
\acro{PLS}{Physical Layer Security}
\acro{SWIPT}{Simultaneous Wireless and Information Power Transfer}
\end{acronym}

\def \E {\mathbb E}

\tikzstyle{int}=[draw, fill=cyan!20, minimum size=2em]
\tikzstyle{int_blue}=[draw, fill=blue!20, minimum size=2em]
\tikzstyle{int_green}=[draw, fill=green!20, minimum size=2em]
\tikzstyle{int_red}=[draw, fill=red!20, minimum size=2em]
\tikzstyle{init} = [pin edge={to-,thin,black}]

\newcommand{\MATLAB}{\textsc{Matlab}}

\newcommand{\ua}{\uparrow}
\newcommand{\nc}{\newcommand}
\nc{\da}{\downarrow} \nc{\hc}{\hat{c}} \nc{\hS}{\hat{S}}
\nc{\bra}{\langle} \nc{\ket}{\rangle} \nc{\eq}{equation (\ref}
\nc{\h}{\hat} \nc{\hT}{\h{T}}\nc{\be}{\begin{eqnarray}}
\nc{\ee}{\end{eqnarray}}\nc{\rd}{\textrm{d}}\nc{\e}{eqnarray}\nc{\hR}{\hat{R}}\nc{\Tr}{\mathrm{Tr}}
\nc{\tS}{\tilde{S}}\nc{\tr}{\mathrm{tr}}\nc{\8}{\infty}\nc{\lgs}{\bra\ua,\phi|}\nc{\rgs}{|\ua,\phi\ket}
\nc{\hU}{\hat{U}}\nc{\lfs}{\bra\phi|}\nc{\rfs}{|\phi\ket}\nc{\hZ}{\hat{Z}}\nc{\hd}{\hat{d}}\nc{\mD}{\mathcal{D}}
\nc{\bd}{\overline{d}}\nc{\bc}{\overline{c}}\nc{\mc}{\mathcal}\nc{\ea}{eqnarray}\nc{\mG}{\mathcal{G}}\nc{\bce}{\begin{center}}
\nc{\ece}{\end{center}}
\date{12th March 2020}

\begin{document}

\title{On the Effect of Correlation on the Capacity of Backscatter Communication Systems}
\author{J.\,L. Matez-Bandera, P. Ramirez-Espinosa, J.\,D. Vega-Sanchez and  F.\,J. Lopez-Martinez
\thanks{Manuscript received March xx, 2020; revised XXX. This work was funded by the Spanish Government (Ministerio de Economia y Competitividad) through grant TEC2017-87913-R, Junta de Andalucia (project P18-RT-3175, TETRA5G)  and the ``Becas Colaboraci\'on con departamentos'' program.}
\thanks{J. L. Matez-Bandera is with Dpto. Ingenieria de Sistemas y Automatica, Universidad de Malaga, 29071 Malaga, Spain. (e-mail: $\rm josematez@uma.es$).}
\thanks{P. Ramirez-Espinosa is with the Connectivity Section, Department  of  Electronic Systems, Aalborg University, Aalborg {\O}st 9220, Denmark. (e-mail: $\rm pres@es.aau.dk$).}
	\thanks{J. D. Vega-Sanchez is with Departamento de Electronica Telecomunicaciones y Redes de Informacion, Facultad de Electrica y Electronica, Escuela Politecnica Nacional (EPN), Quito, Ecuador. (e-mail: $\rm jose.vega01@epn.edu.ec$)}
\thanks{F. J. Lopez-Martinez is with Dpto. Ingenieria de Comunicaciones, Universidad de Malaga, 29071 Malaga, Spain. (e-mail: $\rm fjlopezm@ic.uma.es$)}
\thanks{This work has been submitted for publication. Copyright may be transferred without notice, after which this version may no longer be accessible.}
}

\maketitle
\begin{abstract}
We analyse the effect of correlation between the forward and backward links on the capacity of backscatter communication systems. To that aim, we obtain an analytical expression for the average capacity under a correlated Rayleigh product fading channel, as well as closed-form asymptotic expressions for the high and low signal-to-noise ratio (SNR) regimes. Our results show that correlation is indeed detrimental for a fixed target SNR; contrarily to the common belief, we also see that correlation can be actually beneficial in some instances when a fixed power budget is considered.
\end{abstract}

\begin{IEEEkeywords}
	Backscatter communications, capacity, correlation, fading. 
\end{IEEEkeywords}

\section{Introduction}

Communication using reflected power, usually referred to as backscatter communication, is a rather mature idea \cite{Stockman1948} that is experiencing a tremendous growth in the last decade due to the advent of \ac{RFID} systems and the \ac{IoT}\cite{Griffin2008,Griffin2010,Arnitz2012,Zhang2019}. Because of the low mobility of transmitters, receivers and tags in this context, as well as due to the limited operational range of this technology, there exist a non-negligible correlation between the forward and backward links \cite{Griffin2007,Alhassoun2019}. However, this effect is often overlooked because of the inherent mathematical complexity of the equivalent channel observed by the receiver, which involves a product of correlated random variables.

Most existing results in the literature have investigated the effect of correlation in backscatter communications in terms of the outage probability and bit error rate \cite{Zhang2019,Bekkali2015,Gao2016}, evaluating the performance loss due to correlation for a target \ac{SNR}. The motivation of this work is two-fold: first, we aim to analyze the effect of correlation on the average capacity of a backscatter communication system. Second, we further investigate the impact of correlation ${(a)}$ for a fixed average \ac{SNR} at the receiver ${(b)}$ for a fixed system power budget. Results will show that correlation can be beneficial for system performance in the low \ac{SNR} regime in the latter scenario.

Communication using reflected power, usually referred to as backscatter communication, is a rather mature idea \cite{Stockman1948} that is experiencing a tremendous growth in the last decade due to the advent of \ac{RFID} systems and the \ac{IoT}\cite{Griffin2008,Griffin2010,Arnitz2012,Zhang2019}. Because of the low mobility of transmitters, receivers and tags in this context, as well as due to the limited operational range of this technology, there exist a non-negligible correlation between the forward and backward links \cite{Griffin2007,Alhassoun2019}. However, this effect is often overlooked because of the inherent mathematical complexity of the equivalent channel observed by the receiver, which involves a product of correlated random variables.

Most existing results in the literature have investigated the effect of correlation in backscatter communications in terms of the outage probability and bit error rate \cite{Zhang2019,Bekkali2015,Gao2016}, evaluating the performance loss due to correlation for a target \ac{SNR}. The motivation of this work is two-fold: first, we aim to analyze the effect of correlation on the average capacity of a backscatter communication system. Second, we further investigate the impact of correlation ${(a)}$ for a fixed average \ac{SNR} at the receiver ${(b)}$ for a fixed system power budget. Results will show that correlation can be beneficial for system performance in the low \ac{SNR} regime in the latter scenario.


\section{System Model}

\begin{figure}[ht]
\centering
\begin{tikzpicture}[node distance=3.5cm,auto,>=latex']
    \node [int] (a) {$\rm Tag$};
    \node [int_green, pin={[init]below:$P_{T}$}, left of=a,node distance=3cm] (b) {$\rm{TX}$};
    \node [int_green,pin={[init]below:$P_{R}$}] (c) [right of=a] {$\rm{RX}$};
    \node [coordinate] (end) [right of=c, node distance=2cm]{};
    \path[->] (a) edge node {$h_{\rm b}$} (c);
    \path[->] (b) edge node {$h_{\rm f}$} (a);
\end{tikzpicture}
\caption{Backscatter communication system set-up}
\label{PB_system}
\end{figure}
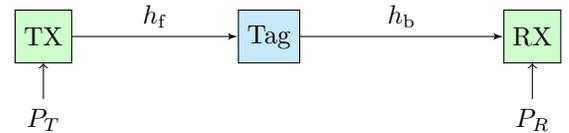

Let us consider a general backscatter communication system consisting on a transmitter, a passive tag and a reader, as depicted in Fig. \ref{PB_system}. 

For simplicity, yet without loss of generality, we assume that the different agents are single-antenna devices. The instantaneous received signal power at the reader can be expressed as \cite{Bekkali2015}
\begin{equation}
\label{EqBekkali}
P_R= P_T L_t  g_f g_b = \overline{P_R}   g_f g_b ,
\end{equation}
where $g_f=\vert h_f\vert^2 $ and $g_b=\vert h_b\vert^2$ indicate the fading power channel coefficients associated to the forward (i.e., transmitter-to-tag) and backward (i.e., tag-to-reader) links. We assume normalised fading coefficients so that $\mathbb{E}\{ g_f\}=\mathbb{E}\{ g_b\}=1$, where $\mathbb{E}\{\cdot\}$ denotes the expectation operator. Hence, the term $\overline{P_R}$ can be regarded as the average receive power when the forward and backward links are independent, while $P_T$ indicates the transmit power. The term $L_t$ encapsulates the effects of polarization losses due to mismatches between antennas, path losses in the forward and backward links, antenna gains, the coding and modulation schemes, and the tag's power transfer efficiency and reflection coefficient. 

As previously indicated, the forward and backward links are in general correlated \cite{Griffin2007,Alhassoun2019} in the context of backscatter communication. The instantaneous \ac{SNR} at the reader will be given by $\gamma= P_R/N_0 $, 
with $N_0$ denoting the noise power.

Similarly to other references in the literature that analyze the effect of correlation in backscatter communications \cite{Griffin2007,He2011}, we consider that the wireless links undergo Rayleigh fading. While other works consider more sophisticated channel models \cite{Gao2016,Zhang2019}, their inherent mathematical complexity prevents from extracting insights for the actual effect of correlation on system performance. With this consideration, the distribution of $\gamma$ is that of the product of two correlated exponential RVs, which can be derived from \cite[eq. 6.55]{Simon2007} as:
\begin{equation}
\label{eqPDF}
f_{\gamma}(\gamma)=\frac{2}{\overline\gamma}\tfrac{1+\rho}{1-\rho}I_0\left(\frac{2}{1-\rho}\sqrt{\frac{\rho\gamma(1+\rho)}{\overline\gamma}}\right)K_0\left(\frac{2}{1-\rho}\sqrt{\frac{\gamma(1+\rho)}{\overline\gamma}}\right),
\end{equation}
where $\overline\gamma=\mathbb E\left\{\frac{P_R}{N_0}\right\}$ denotes the average SNR at the receiver side, the power correlation coefficient is defined as $\rho\triangleq\frac{{\rm cov}\{g_f g_b\}}{\sqrt{{\rm var}\{g_f\}{\rm var}\{g_b\}}}$, $I_0(\cdot)$ is the modified Bessel function of the first kind and order zero, and $K_0(\cdot)$ is the modified Bessel function of the second kind and order zero.

Manipulating \eqref{EqBekkali}, we can re-express the instantaneous SNR as
\begin{align}
	\gamma= &\, \underbrace{\frac {P_{T} L_t }{N_0}}_{\overline\gamma^I=P_T/N_E} g_f g_b.
\end{align}
For convenience of discussion, we define the parameter $\overline\gamma^I$ as the ratio between the system transmit power and a constant term $N_E$, which can be regarded as the system's noise referred to the transmitter output. We note that $\overline\gamma^I$ reduces to the average SNR at the receiver side \emph{only} in the absence of correlation. In such case, we have $\mathbb{E}\{g_f g_b\}=\mathbb{E}\{g_f\}\mathbb{E}\{g_b\}=1$ because of the definition of normalised channel gains. In the general case of correlation between the forward and backward links, from \cite[eq. 6.55]{Simon2007} we have that $\mathbb{E}\{g_f g_b \}=1+\rho$. A first important remark is in order at this point: for a fixed transmit power $P_T$, the average SNR at the receiver end is increased because of correlation, i.e., $\mathbb{E}\{\gamma\}=\overline\gamma^I(1+\rho)$. While this seems beneficial from a system design perspective, it turns out that it also increases the variance of $\gamma$. In the following section, we aim to determine the impact of both effects on system capacity.

\section{Effect of correlation on system performance}

The average capacity per bandwidth unit is defined as
\begin{equation}
\label{eqC}
\overline{C}[\text{bps/Hz}]\triangleq\int_0^{\infty}\log_2(1+\gamma)f_\gamma(\gamma)d\gamma,
\end{equation}
and can be numerically evaluated plugging \eqref{eqPDF} into \eqref{eqC} and using standard integration routines included in commercial software packages (e.g., \texttt{integral} in \MATLAB, and the scaled in-built representations of the modified Bessel functions). However, it is also possible to find an expression in terms of well-known special functions conventionally used for capacity analyses. After some manipulations, \eqref{eqC} reads as
\begin{align}\label{Capacity}
\overline{C}={C_1\int_0^{\infty}
     \log(1+\gamma)I_0\left(b\sqrt{\gamma}\right)K_0\left(a\sqrt{\gamma}\right)d\gamma,}
\end{align}
{where $\log(\cdot)$ denotes natural logarithm, $a=\tfrac{2}{1-\rho}\sqrt{\tfrac{(1+\rho)}{\overline\gamma}}$, $b=a \sqrt{\rho}$, and $C_1=\tfrac{a^{2}(1-\rho)}{2\log\left ( 2 \right )}$. }
{In order to find an analytical solution for~\eqref{Capacity}, we resort to a series expansion of $I_0(\cdot)$~\cite[id.~(03.02.02.0001.01)]{Wolfram} and the Meijer's G-function $G_{p, q}^{m, n}\left[\cdot
\right]$ representations of both $K_0(\cdot)$~\cite[id.~(03.04.26.0008.01)]{Wolfram} and $\log(\cdot)$~\cite[id.~(01.04.26.0003.01)]{Wolfram}. Thus,~\eqref{Capacity} can be rewritten as}
\begin{align}\label{Capacity2}
{
\overline{C}= }&{C_1 \sum_{k=0}^{\infty} C_k\underset{I_1}{\underbrace{ \int_{0}^{\infty} \gamma^{k} G_{2,2}^{1,2}\left[ \gamma \bigg|
\begin{array}{c}
 1,1 \\
 1,0 \\
\end{array}
\right] G_{0,2}^{2,0}\left[ \frac{a^2\gamma}{4} \bigg|
\begin{array}{c}
 - \\
 0,0 \\
\end{array}
\right] d\gamma}}}
\end{align}
{where  $C_k=\frac{1}{2}\tfrac{1}{k!\Gamma(k+1)}\left ( \tfrac{b}{2} \right )^{2k}$.}
{Finally, by solving $I_1$ with the help of~\cite[id.~(07.34.21.0012.01)]{Wolfram},
the expression for the average capacity can be formulated as}
\begin{align}\label{Capacity3}
{
\overline{C}= }&{C_1  \sum_{k=0}^{\infty} C_k \hspace{0.5mm} \mathrm{H}_{2,4}^{4,1}\left[\frac{a^2}{4}\bigg|
\begin{array}{c}
 (-k-1,1),\hspace{4.5mm} (-k,1) \\
 (0,1), (0,1), (-k-1,1),(-k-1,1)\\
\end{array}
\right],} 
\end{align}
{where $\mathrm{H}_{p,q}^{m,n}\left[ \cdot
\right]$ is the Fox H-function~\cite[Eq.~(1.1)]{Fox}}. We note that several efficient implementations of the H-function are readily available in the literature; in our case, we use the one reported in~\cite{Vega} to evaluate our results. Now, in order to gain insights into the effect of correlation on capacity, we resort to asymptotic analysis in the high and low SNR regimes.

In the high SNR regime, the average capacity can be approximated as in \cite{Yilmaz2012}
\begin{equation}
\label{aAC}
\overline{C}(\overline\gamma)_{\overline\gamma\Uparrow}\approx \log_2(\overline\gamma) + \log_2(e)\left.\frac{d\mathcal{M}(k)}{dk}\right\rvert_{k=0},
\end{equation}
where $M(k)$ are the normalised moments of $\gamma$, which can be expressed from \eqref{eqPDF} as
\begin{align}
\label{mom}
\mathcal{M}(k)\triangleq
\frac{\E\{\gamma^k\}}{\overline\gamma^k}= (1-\rho)^{-k}\Gamma(1+k)^2{}_2F_1\left(-k,-k; 1; \rho\right),
\end{align}
where $\Gamma(\cdot)$ is the gamma function and ${}_2F_1(\cdot)$ denotes the Gauss hypergeometric series \cite[eq. (15.1.1)]{abramowitz1964}.

The derivative of \eqref{mom} with respect to $k$ is calculated by repeatedly applying the derivative chain rule. Taking into account that 
\begin{equation}
	\label{eq:DerGamma}
	\frac{d}{dn}\Gamma(a) = \Gamma(a)\psi(a)
\end{equation}
with $\psi(\cdot)$ the digamma function, we only need to get an expression for the partial derivative of the Gauss series. In order to do so, we calculate
\begin{equation}
	\frac{{\partial^2}}{\partial a \partial b}{}_2F_1(-a,-b;1;\rho)
\end{equation}
where $a,b > 0$ by expressing the hypergeometric function in its series form as
\begin{equation}
	{}_2F_1(-a,-b;1;\rho) = \sum_{m=0}^\infty\frac{(-a)_m(-b)_m}{(1)_m}\frac{\rho^m}{m!}
\end{equation}
with $(\cdot)_m$ denoting the Pochhammer symbol. Then, using \cite[p. 17 eq. (9)]{Srivastava1985} we obtain
\begin{equation}
	\frac{{\partial^2}}{\partial a \partial b}{}_2F_1(-a,-b;1;\rho) = \frac{{\partial^2}}{\partial a \partial b} \sum_{m=0}^{\min(a,b)} \tfrac{\Gamma(a+1)\Gamma(b+1)}{\Gamma(a-m+1)\Gamma(b-m+1)}\frac{\rho^m}{(m!)^2}. 
\end{equation}
Finally, the application of \eqref{eq:DerGamma} and the chain rule lead us to
\begin{align}
	\frac{{\partial^2}}{\partial a \partial b}{}_2F_1(-a,-b;1;\rho) &= \sum_{m=0}^{\min(a,b)} \tfrac{\Gamma(a+1)\left[\psi(a+1)-\psi(a-m+1)\right]}{\Gamma(a-m+1)} \notag \\
	&\times\tfrac{\Gamma(b+1)\left[\psi(b+1)-\psi(b-m+1)\right]}{\Gamma(b-m+1)}\frac{\rho^m}{(m!)^2}.
\end{align} 
Using the above result, and after some algebraic manipulations, the following expression for \eqref{aAC} holds:
\begin{equation}
\label{aAC2}
\overline{C}(\overline\gamma)_{\overline\gamma\Uparrow}\approx \log_2(\overline\gamma) - 2\log_2(e)\gamma_e -\log_2(1+\rho),
\end{equation}
where $\gamma_e=0.57721\ldots$ is the Euler-Mascheroni constant. Inspection of \eqref{aAC2} yields important insights from a practical perspective. First, we see that for a fixed SNR at the receiver, correlation is always detrimental for capacity. We also see that in the absence of correlation, the capacity loss term is doubled with respect to the non-backscattered Rayleigh case, for which $\overline{C}\approx \log_2(\overline\gamma) - \log_2(e)\gamma_e$.

Let us now analyze the effect of correlation for a \emph{fixed transmit power}, i.e., for a given $P_T$. In this scenario, we have that $\overline\gamma=P_T/N_E(1+\rho)=\overline\gamma^I(1+\rho)$, so that \eqref{aAC2} reduces to
\begin{equation}
\label{aAC3}
\overline{C}(\overline\gamma^I)_{\overline\gamma^I\Uparrow}\approx \log_2(\overline\gamma^I) - 2\log_2(e)\gamma_e.
\end{equation}
Strikingly, in the high SNR regime the degradation due to correlation is compensated by the increase in average SNR compared to the case of independent forward and backward links. This is a novel observation in the literature to the best of the authors' knowledge. Conventionally, the widespread understanding is that correlation is always detrimental for system performance. As we have now seen, this is only the case when considering a fixed receive SNR, not for a fixed transmit power budget.

As the SNR is decreased, we have that $\log_2(1+\gamma)\approx\gamma$; hence, capacity in the low SNR regime is approximated by the first moment of the SNR. We now see that correlation is actually beneficial for system operation when a fixed $P_T$ is considered, i.e.,
\begin{equation}
\label{aAC4}
\overline{C}(\overline\gamma)_{\overline\gamma\Downarrow}\approx \overline\gamma = \overline\gamma^I(1+\rho).
\end{equation}

\section{Numerical Results}

\begin{figure}[t]
	\centering
     \includegraphics[width=.99\columnwidth]{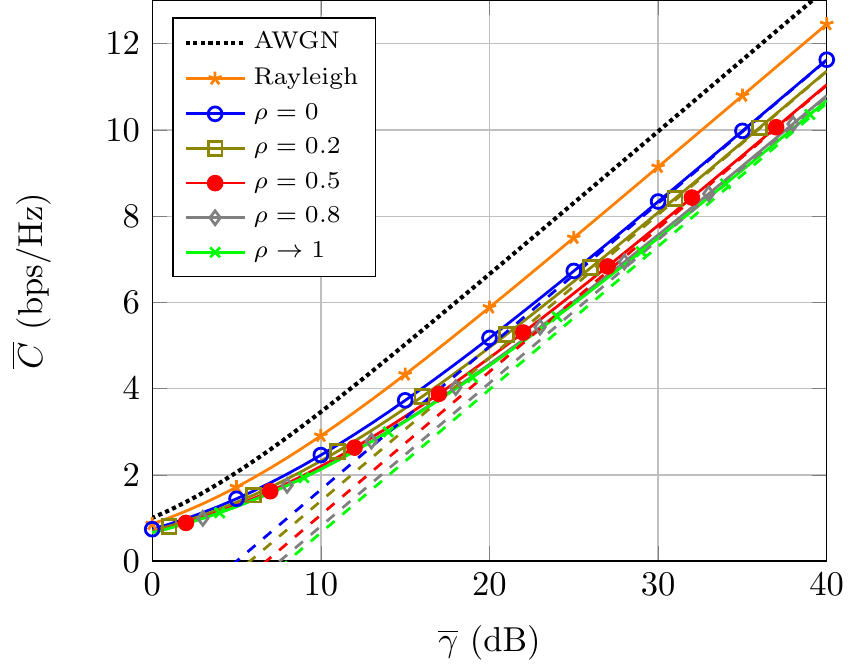}
         \caption{Average capacity $\overline{C}$ as a function of the average SNR at the receiver $\overline\gamma$, for different values of the power correlation coefficient $\rho$. Dashed lines correspond to the asymptotic results using \eqref{aAC2}. Markers correspond to MC simulations.}
      \label{fig:1}
\end{figure}

We evaluate the average capacity for the investigated scenario, in order to determine the effect of channel correlation on system performance. In order to double-check the validity of our expressions, we also include Monte Carlo (MC) simulations on the top of our theoretical results. The cases of non-backscattered Rayleigh fading and no fading (AWGN) are also included as reference values. In Fig. \ref{fig:1} we evaluate the average capacity for a fixed average SNR at the receiver side, i.e., a fixed value of $\overline\gamma$. We see that as correlation is increased, the capacity loss with respect to the case of independent backscatter is increased. In the limit case of total correlation, the capacity loss in the high-SNR regime is about $1$ bps/Hz larger than for $\rho=0$. We also observe that the asymptotic capacity results become tight as $\overline\gamma$ grows. In all instances, as predicted by Jensen's inequality, capacity is always lower than in the AWGN case -- and also lower than in the single Rayleigh case. We note that despite correlation degrades system performance for a fixed $\overline\gamma$, the transmit power requirements are reduced by a factor of $1+\rho$ compared to the case of independent fading.

\begin{figure}[t]
	\centering
     \includegraphics[width=.99\columnwidth]{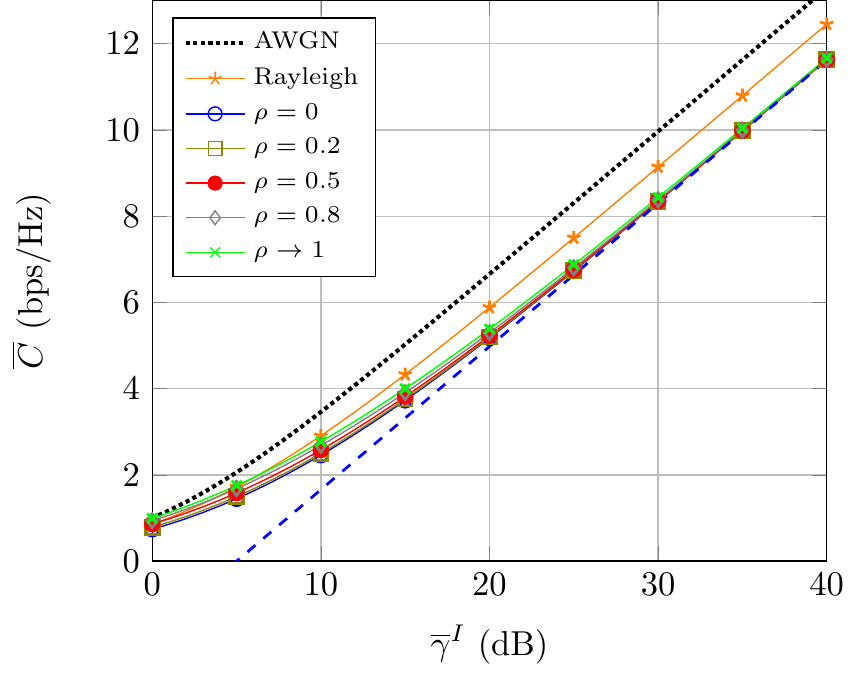}
         \caption{Average capacity $\overline{C}$ as a function of the average SNR at the receiver in the absence of correlation $\overline\gamma^I$ (i.e., for fixed transmit power $P_T$), for different values of the power correlation coefficient $\rho$. Dashed lines correspond to the asymptotic results using \eqref{aAC3}. Markers correspond to MC simulations.}
      \label{fig:2}
\end{figure}

\begin{figure}[t]
	\centering
     \includegraphics[width=.99\columnwidth]{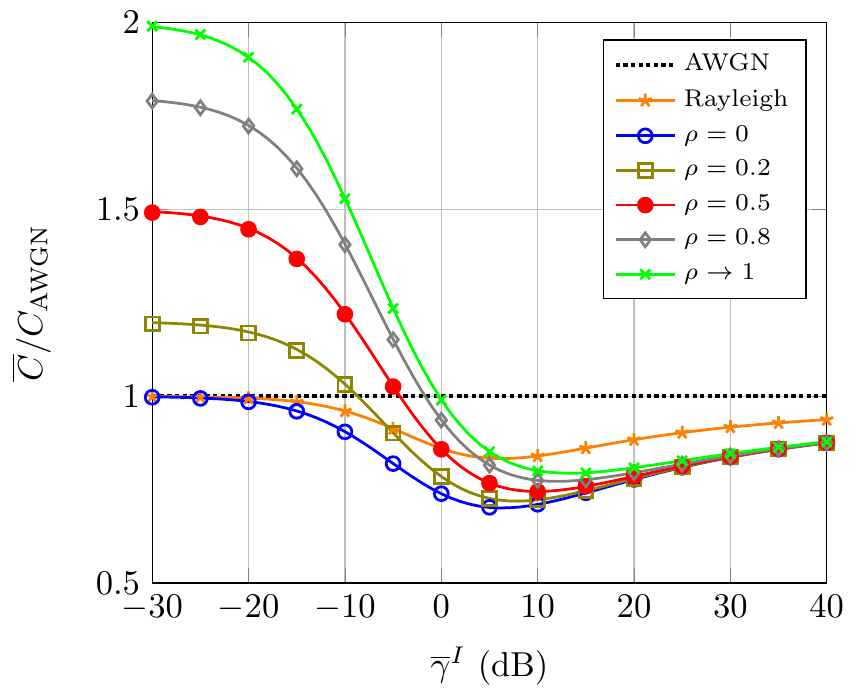}
         \caption{Average capacity $\overline{C}$ normalised to that of the AWGN case, as a function of the average SNR at the receiver in the absence of correlation $\overline\gamma^I$ (i.e., for fixed transmit power $P_T$), for different values of the power correlation coefficient $\rho$. Markers correspond to MC simulations. For very low SNR, capacity tends to the asymptotic value in \eqref{aAC4}.}
      \label{fig:3}
\end{figure}

Fig. \ref{fig:2} now evaluates the case of using a fixed transmit power $P_T$. In the absence of correlation, the average SNR corresponds to that denoted by $\overline\gamma^I$. As previously discussed, as correlation is increased then the average SNR experienced at the receiver side is also increased by a factor $(1+\rho)$. We see that for a fixed system power budget, the effect of correlation is vanished in the high-SNR regime. The capacity degradation due to correlation is compensated by the increase in average SNR and hence the effect of $\rho$ becomes immaterial as the average SNR grows, as predicted by the asymptotic results in \eqref{aAC3}. We also observe that as the average SNR is reduced, correlation is actually \emph{beneficial} for system capacity -- this is a novel observation in the literature to the best of our knowledge. In order to better illustrate the role of correlation in the low SNR regime, we represent in Fig. \ref{fig:3} the average capacity normalised to that of the AWGN case.

We see in Fig. \ref{fig:3} that as the transmit power is decreased (i.e., which causes the average SNR to be reduced for a given system set-up), correlation allows for obtaining a larger capacity than in the case of independent backscattering, and remarkably, also than in the absence of fading. Note that this result does not contradict Jensen's inequality, as it gives an upper bound for channel capacity for a \emph{fixed} average receive SNR. Instead, it suggests that for a fixed power budget and a low SNR operation, correlation can be beneficial for system performance.

\section{Conclusion}
We provided relevant conclusions about the role of channel correlation in the context of backscatter communications. While our analytical results confirm that correlation is detrimental for capacity for a target average SNR, they also show that when a fixed system power budget is considered, correlation is actually beneficial in the low SNR regime and its impact on system performance vanishes as the average SNR is increased.
%
%
%

%
%
\bibliographystyle{ieeetr}
\bibliography{cbs.bib}

\end{document}